\documentclass[prb,showpacs,twocolumn,amsmath,amssymb,groupadress,superscriptaddress]{revtex4}
\usepackage{graphicx}
\usepackage{dcolumn}
\usepackage{bm}
\usepackage{epstopdf}
\usepackage{srcltx}
 
\begin{document}

\preprint{APS/PRB -Khalyavin}

\title{Long-range magnetic order in CeRu$_2$Al$_{10}$ evidenced by $\mu^+$SR and neutron diffraction}

\author{D.D. Khalyavin}
\affiliation{ISIS facility, Rutherford Appleton Laboratory-STFC,
Chilton, Didcot, Oxfordshire, OX11 0QX, UK }
\author{A.D. Hillier}
\affiliation{ISIS facility, Rutherford Appleton Laboratory-STFC,
Chilton, Didcot, Oxfordshire, OX11 0QX, UK }
\author{D.T. Adroja}
\affiliation{ISIS facility, Rutherford Appleton Laboratory-STFC,
Chilton, Didcot, Oxfordshire, OX11 0QX, UK }
\author{A.M. Strydom}
\affiliation{Physics Department, University of Johannesburg, PO Box 524, Auckland Park 2006, South Africa }
\author{P. Manuel}
\affiliation{ISIS facility, Rutherford Appleton Laboratory-STFC,
Chilton, Didcot, Oxfordshire, OX11 0QX, UK }
\author{L.C. Chapon}
\affiliation{ISIS facility, Rutherford Appleton Laboratory-STFC,
Chilton, Didcot, Oxfordshire, OX11 0QX, UK }
\author{P. Peratheepan}
\affiliation{Physics Department, University of Johannesburg, PO Box 524, Auckland Park 2006, South Africa }
\author{K. Knight}
\affiliation{ISIS facility, Rutherford Appleton Laboratory-STFC,
Chilton, Didcot, Oxfordshire, OX11 0QX, UK }
\author{P. Deen}
\affiliation{Institute Laue-Langevin, 6 Rue Jules Horowitz, BP 156, 38042 Grenoble Cedex 9, France }
\author{C. Ritter}
\affiliation{Institute Laue-Langevin, 6 Rue Jules Horowitz, BP 156, 38042 Grenoble Cedex 9, France }
\author{Y. Muro}
\affiliation{Department of Quantum matter, ADSM, and IAMR, Hiroshima University, Higashi-Hiroshima, 739-8530, Japan }
\author{T. Takabatake}
\affiliation{Department of Quantum matter, ADSM, and IAMR, Hiroshima University, Higashi-Hiroshima, 739-8530, Japan }
\date{\today}

\begin{abstract}
The low temperature state of CeRu$_2$Al$_{10}$ has been studied by neutron powder diffraction and muon spin relaxation ($\mu^{+}$SR). By combining both techniques, we prove that the transition occurring below T$^*\sim$27K, which has been the subject of considerable debate, is unambiguously magnetic due to the ordering of the Ce sublattice. The magnetic structure with propagation vector {\bf k}=(1,0,0) involves collinear antiferromagnetic alignment of the Ce moments along the $c$-axis of the $Cmcm$ space group with a reduced moment of 0.34(2)$\mu_B$. No structural changes within the resolution limit have been detected below the transition temperature. However, the temperature dependence of the magnetic Bragg peaks and the muon precession frequency show an anomaly around T$_{2}\sim$12K indicating a possible second transition. 
\end{abstract}

\maketitle

\begin{figure}[t]
\includegraphics[scale=0.80]{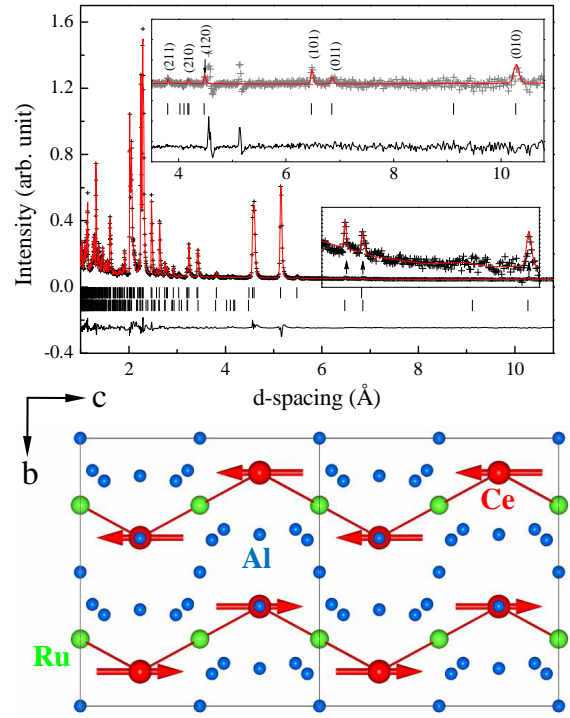}
\caption{(Color online) Rietveld refinement of neutron powder diffraction pattern obtained at 1.5K. The cross symbols and (red) solid line represent the experimental and calculated intensities, respectively, and the line below is the difference between them. Tick marks indicate the positions of Bragg peaks for the nuclear scattering ($Cmcm$ space group, top line) and magnetic scattering ({\bf k}=(1,0,0) propagation vector, bottom line). The inset shows the refinement of the pure magnetic scattering, obtained by taking the difference between data sets at 1.5K and 30K. The asymmetric features are due to the difference in positions of the nuclear peaks at different temperatures (thermal expansion). Bottom panel: schematic representation of the magnetic structure of CeRu$_2$Al$_{10}$, projected into ($bc$) plane}
\label{fig:rietveld}
\end{figure}

\indent A novel phase transition recently discovered in the orthorhombic heavy fermion compound CeRu$_2$Al$_{10}$ has attracted great attention due to the possible unconventional physics behind it.\cite{ISI:000271357300025} Initially, the transition was attributed to magnetic ordering of the cerium moments \cite{ISI:000271357300025}, based on macroscopic measurements which revealed pronounced anomalies in various physical properties. However, the unexpectedly high critical temperature (T$^*\sim$27K) and magnetic susceptibility data obtained for different crystallographic directions,\cite{ISI:000272985600010} as well as $^{27}$Al NQR/NMR measurements \cite{ISI:000272985600018} raised doubts about the magnetic nature of this transition. Several alternative interpretations have recently been proposed \cite{ISI:000272985600010,ISI:000272985600018,ISI:000276688700010}, but the most popular scenario involves the formation of a spin-singlet state of Ce suggested by Tanida et al.\cite{ISI:000276688700010} This would corresponds to an effective reduction of dimensionality of the system due to structural distortions, namely a spin-Peierls transition.\cite{Hanzawa_1} This idea was backed-up by recent neutron scattering experiments revealing a possible superstructure below T$^*$ and the presence of low-energy modes interpreted as singlet-triplet excitations from dimerized Ce ions.\cite{arXiv:1003.4933v2} The origin of the superstructure reflections was however not clarified experimentally and only a recent theoretical work by Hanzawa\cite{Hanzawa_2} attributed them to displacements of some Al sites with an associated symmetry-lowering from $Cmcm$ down to $Pmnn$. Understanding the origin of the superstructure reflections is obviously key to determining the ground state of CeRu$_2$Al$_{10}$, and should allow for a quantitative analysis of the order parameter at the phase transition. \\
\indent In the present communication, we show by combining the results of $\mu^+$SR and powder neutron diffraction that the transition at T$^*$ is magnetic in origin and involve long-range order of the Ce moments. Below T$^*$, the observation in the neutron data of several superlattice reflections not detected previously  \cite{arXiv:1003.4933v2} indexed by the propagation vector k=(1,0,0), enables a full analysis of the magnetic configuration. Refinement of the data shows that the best model corresponds to an antiferromagnetic order of the Ce moment along the $c$-axis of the $Cmcm$ space group. The data can not be explained by magnetic ordering of the Ru sublattice or by a structural transition, and is uniquely compatible with one of the symmetry modes calculated for the Ce site. The reduction of paramagnetic background below T$^*$ in the neutron data and the independent observation of coherent oscillations due to internal field in the $\mu^+$SR, confirms the magnetic origin of the transition. Anomalies detect primarily in the $\mu^+$SR data indicate that additional transitions, which origins are not yet understood, appear below T$^*$. \\
\indent A polycrystalline sample of CeRu$_2$Al$_{10}$ used in the powder neutron diffraction and zero-field muon spin relaxation experiments was prepared by ultra-high purity argon arc melting of stoichiometric quantities of the starting elements. Both experiments were carried out at the ISIS pulsed neutron source of the Rutherford Appleton Laboratory, U.K, respectively on the MuSR spectrometer and WISH diffractometer on the second target station (TS-2). The $\mu^+$SR experiments were conducted in longitudinal geometry. The powdered sample was mounted onto a 99.995+\% pure silver plate. The sample and mount were then inserted into a cryostat with a temperature range of 1.2 K to 300 K. Any Ag exposed to the muon beam would give a flat time independent background. For the neutron diffraction experiments, a 6 g sample was loaded into a cylindrical 6 mm vanadium can, placed in an Oxford Instrument cryostat. Data were recorded in the temperature interval 1.5K - 30K, with long counting times (8 hours) at T=1.5K and T=30K. Intermediate temperature points were measured with a lower exposition time ($\sim$2h). The program FullProf\cite{ISI:A1993ME99200007} was used for Rietveld refinements and group-theoretical calculations were performed with the aid of the ISOTROPY software.\cite{ISOTROPY}\\
\begin{figure}[t]
\includegraphics[scale=0.80]{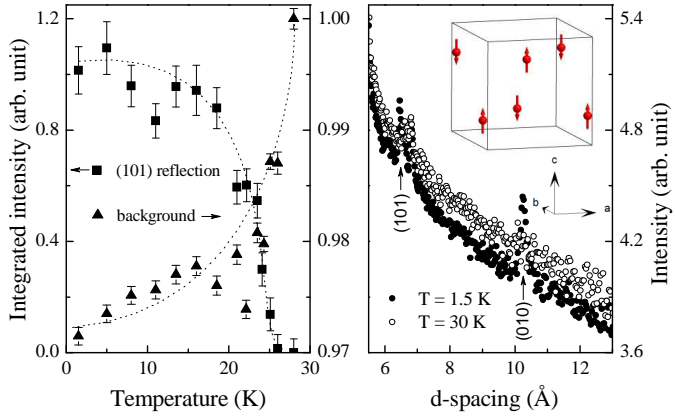}
\caption{(Color online) Integrated intensity of the (101) reflection and the background as a function of temperature (left panel). A low-Q region of two diffraction patterns collected at T=1.5K and T=30K (right panel). In both cases, the dot lines are a guide for eyes. The inset shows ordering of Ce magnetic moments at T=1.5K.}
\label{fig:Integr_inten}
\end{figure}
\indent The neutron diffraction patterns collected at T$>$T$^*$ are consistent with the $Cmcm$ symmetry and can be satisfactorily fitted (R$_{Brag}$=4.95\%) in the structural model proposed by Thiede, Ebel and Jeitschko\cite{ISI:000071595100024}with the unit cell parameters at T=30K being: $a$=9.1322(2)$\AA$, $b$=10.2906(2)$\AA$ and $c$=9.1948(2)$\AA$. Below T$^*$, a set of new reflections (Fig. \ref{fig:Integr_inten}) associated with the {\bf k}=(1,0,0) propagation vector (Y-point of symmetry in Miller and Love notations \cite{rep:MillerLove}) appears indicating the phase transition detected in previous investigations by numerous experimental techniques.\cite{ISI:000271357300025,ISI:000272985600010,ISI:000272985600018,ISI:000276688700010} These reflections are clearly visible only at low momentum-transfer(Q), whilst there are no detectable changes in the region at high-Q. By taking the difference between the patterns collected at T=1.5K and T=30K, seven peaks whose intensities are significantly higher than the error bars are detected at positions in $d$-spacing $>$3$\AA$ (Fig. \ref{fig:rietveld}). Another important feature is the reduction of background observed at low-Q and associated with the appearance of the superlattice reflections below T$^*$ (Fig. \ref{fig:Integr_inten}). This is consistent with the expected suppression of paramagnetic scattering and the transfer of intensity to Bragg scattering in the event of a magnetic phase transition. This is best seen in the temperature dependence of the integrated intensity of the strongest superlattice peak (101) and the background integrated in the range 7$\AA$-10$\AA$ shown in Fig. \ref{fig:Integr_inten} (left panel).\\
\indent To obtain an appropriate model for the magnetic structure, we employed a method whereby combinations of axial vectors localized on the 4$c$(Ce) and 8$d$(Ru) sites and transforming as basis functions of the irreducible representations (irreps) of the wave-vector group ({\bf k}=(1,0,0)), are systematically tested. \cite{rep:Izumov} In agreement with the Landau theory of continuous transition, we found that a single irrep is involved. A unique solution (R$_{nucl}$=4.72\% R$_{mag}$=14.65\%) associated with the Y$^-_3$ irrep was found assuming magnetic ordering of the Ce sublattice (Fig. \ref{fig:rietveld}). None of the magnetic modes spanned by a single-irrep for the 8$d$(Ru) sites was able to fit the data properly. This is consistent with the absence of transition in the isostructural compound LaRu$_2$Al$_{10}$.\cite{ISI:000276688700010,ISI:000071595100024} The obtained model consists of collinear antiferromagnetic ordering of the Ce moments aligned along the $c$-axis of the $Cmcm$ space group (Fig. \ref{fig:Integr_inten} (inset) and \ref{fig:rietveld} (bottom)). The antiparallel configuration is adopted by the ions related by the glide plane and the centering lattice translation. The latter comes directly from the propagation vector which in respect of the primitive unit cell is {\bf k}=(1/2,1/2,0). The corresponding magnetic space group is $C$$_p$$m'cm$ (number 63.11.521 in the Litvin's classification scheme).\cite{ISI:000255049700007}
\begin{figure}[t]
\includegraphics[scale=1.00]{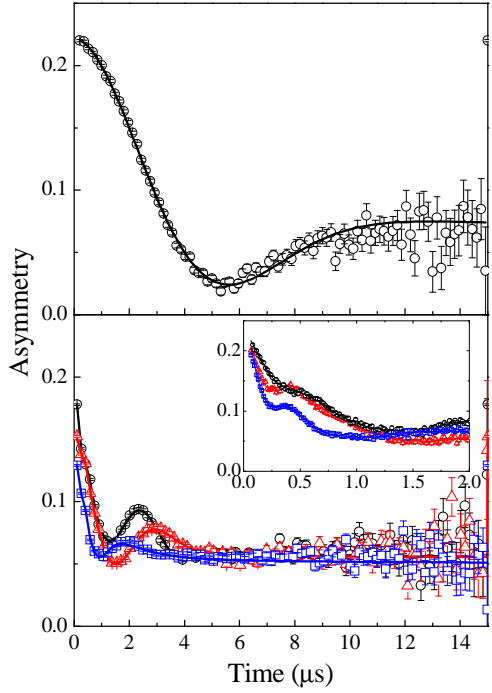}
\caption{(Color online) The upper graph shows the muon depolarisation spectra at 30K. The line is a least squares fit to the data as using Eqn. \ref{eq:fitfun2}. The lower graph shows the muon depolarisation spectra for a range of temperatures (namely 1.4K blue, 10.8K red and 23.0K black). The insert shows the same data but at shorter times. In both cases the line is a least squares fit to the data using Eqn. \ref{eq:fitfun3}.}
\label{fig:Long_T_dep}
\end{figure}
The refined value of the Ce magnetic moment is 0.34(2) $\mu_B$ at T=1.5K and its temperature dependence varies as the square root of the integrated intensity of the (101) reflection (Fig. \ref{fig:Integr_inten}). A small anomaly around T$_2\sim$12K possibly indicates an additional transition, but data on single crystal would be necessary to determine the exact nature of this transition. We note that this anomaly is also observed in the $\mu^+$SR data discussed below.\\
\begin{figure}[t]
\includegraphics[scale=1.00]{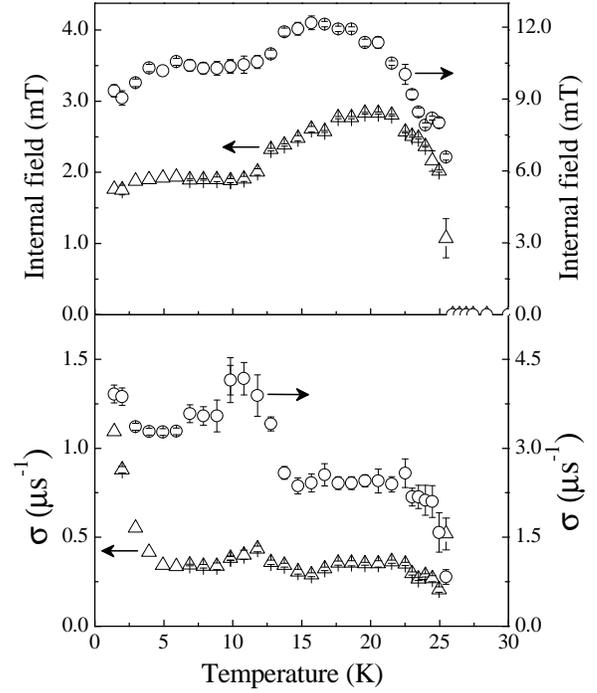}
\caption{The upper graph shows the temperature dependence of the internal field at the muon sites. The  $\triangle$ are associated with the left axis and the $\bigcirc$ are associated with the right axis. The lower graph shows the temperature dependence of the muon depolarisation rate and again the $\triangle$ left axis and the $\bigcirc$ are associated with the right axis.}
\label{fig:muonresults}
\end{figure}
\indent An alternative explanation for the transition below T$^*$, which is widely discussed in the literature, is the dimerization of Ce, i.e. a spin-Peierls transition where the formation of a spin-gapped singlet state drives a structural modulation. Several space groups were proposed based on different experiments\cite{ISI:000272985600018} and theoretical considerations.\cite{Hanzawa_2} Our analysis does not support this model. First of all, qualitatively, one would expect in such a case that the superstructure reflections in  neutron diffraction would be observed at high momentum transfer since their intensity would not be modulated by the magnetic form factor. In addition, at such low temperatures, the intensity is not damped by high thermal parameters. Secondly, it appears that a structural distortion can be discarded by a quantitative analysis: the only isotropy subgroup associated with the Y symmetry point that can account for all observed superstructure reflections is $Pmmn$. An unconstrained refinement with this symmetry was unstable and therefore another approach based on symmetry-adapted atomic displacements was employed. Considering all atoms in the unit-cell (Ce,Ru,Al), there are eleven modes with the Y$^-_4$ symmetry that can act as primary order parameters at the $Cmcm$ $\rightarrow$ $Pmmn$ displacive phase transition. All of these modes were introduced in the refinement either independently or combined but none could give a significant contribution to the observed superstructure peaks, confirming further the magnetic origin. It should be pointed out however that this result does not imply that the symmetry of the low temperature phase is necessarily $Cmcm$, nor does it preclude the presence of very weak secondary structural modes beyond the sensitivity of our measurements.\\
\indent Further arguments in favor of magnetic long-range order is obtained from $\mu^+$SR measurements.
The muon spin relaxation for a metallic system that does not exhibit magnetic order will only be sensitive to the local fields associated with the nuclear spins. Normally, on the time scale of the muon, these nuclear spins are static and randomly orientated. Therefore, the $\mu^+$SR spectra can be described by the Kubo-Toyabe function\cite{ISI:A1979HK06900002}
\begin{equation}
\label{eq:fitfun}
G_z^{KT}(t)=(\frac{1}{3}+\frac{2}{3}(1-\Delta ^2 t^2)\exp(-\frac{\Delta^2t^2}{2})),
\end{equation}
\noindent where $\Delta/\gamma_\mu$ is the local field distribution width and  $\gamma_\mu / 2\pi$=135.5MHz/T is the muon gyromagnetic ratio. In the magnetically ordered state, the muon polarisation will precess with a frequency which is  directly proportional to the ordered moment. The spectra above the phase transition temperature (T$^{*}\sim$27K) are best described (see Fig. \ref{fig:Long_T_dep} (Top)) by 
\begin{equation}
\label{eq:fitfun2}
G_z(t)=A_{0}G_z^{KT}(t)\exp(-\lambda t) + A_{bckgrd}
\end{equation}
\noindent where $\lambda$ is the relaxation rate associated with the dynamic electronic spin fluctuations. As the transition at $\sim$27K is approached, from above, there is a slight increase in the $\lambda$ term whereas $\Delta$ remains temperature independent ($\Delta$=0.30~$\mu s^{-1}$). This increase in $\lambda$ is indicative of approaching a magnetic transition. Now as the temperature is decreased still further the $\mu^+$SR spectra clearly show the presence of coherent oscillations. The presence of these oscillations with one frequency have been previously reported,\cite{Kambe} however, our analysis shows that there are at least two precession frequencies, which is consistent with our $\mu^+$SR results on CeOs$_{2}$Al$_{10}$\cite{adroja10} (see Fig. \ref{fig:muonresults} (bottom and insert)) and indeed our spectra are well described using the following equation,
\begin{eqnarray*}
\label{eq:fitfun3}
G_{z}(t)&=&\sum_{i=1}^2 A_{o,i} exp(-\sigma_i ^2t^2/2)cos(\gamma_\mu B_i t + \varphi) \\ \nonumber
&&+A_{r,i}exp(-\lambda_{i}t)+A_{bck} 
\end{eqnarray*}
\noindent where $B_i$ is the internal field,  $\sigma$ is the depolarisation rate, $\lambda$ is, again, the spin fluctuations along the $z$ direction, $\varphi$ is the phase of the oscillations and finally A$_{bck}$ is the background. The value of A$_{bck}$ was determined from the spectra above the transition temperature and fixed for all the fits below. Fig. \ref{fig:muonresults} shows the temperature dependence of the internal field and the depolarisation rate. These results show that the presence of the procession in the $\mu^+$SR spectra occurs at $\sim$27K. If we examine the amplitudes of the oscillating term then we can conclude that the sample is fully magnetically ordered. A rapid increase in both frequencies  has been observed below T$^*\sim$27K, in addition a broad hump around 20K, which then decreases and plateaus at 12K. Further decreasing the temperature below 12K, a decrease in the internal field has been observed below 5K. Now considering the depolarisation rate, we can see that there is again a broad hump close to 20K which coincides with the broad hump in the internal field at about the same temperature. Interestingly, there is a peak in $\sigma$ at T$_2\sim$12K. This, also, coincides with the anomaly in the temperature dependence of the integrated intensity of the (101) magnetic peak (see Fig. \ref{fig:Integr_inten}) and possibly indicates an additional phase transition. \\
\indent Thus, the $\mu^+$SR data and diffraction measurements leave no doubts about the presence of the long-range magnetic ordering in CeRu$_2$Al$_{10}$. The origin of the high critical temperature and low value of the ordered moment is not clear at present and require further experimental and theoretical efforts. The latter, can be possibly related to a particularly strong hybridization between Ce 4$f$ and conduction electrons. The high ordering temperature is difficult to attribute to RKKY interactions only. In this respect, an important observation is the lack of transition in the isostructural compound CeFe$_2$Al$_{10}$.\cite{ISI:000071595100024} This emphasizes the significant role played by the Ru sublattice in transmitting interactions between the well-separated Ce ions (Fig. \ref{fig:rietveld} (bottom)). The main difference between Fe and Ru is the diffuseness of the outer electronic shells (5$p$/5$d$ versus 4$p$/4$d$) whose effect will need to be investigated by $ab$-initio calculations in the light of our experimental finding.\\
\indent In conclusion, the orthorhombic compound CeRu$_2$Al$_{10}$ exhibits a phase transition at T$^*\sim$27K associated with long-range magnetic ordering of the Ce sublattice. The propagation vector of the ordered state is {\bf k}=(1,0,0) and does not change with temperature. The magnetic structure at T=1.5K involves a collinear antiferromagnetic orientation of the Ce moments along the $c$-axis of the $Cmcm$ space group with a magnitude of 0.34(2)$\mu_B$. The small value of the magnetic moments is not related to a presence of a disordered component and is an intrinsic property of the fully ordered state. Around T$_2\sim$12K a possible structural/magnetic phase transition takes place as seen through an anomalous behaviour of the muon depolarization rate and procession frequencies.

\end{document}